\documentclass[11pt,twoside]{article}
\usepackage{asp2004}
\usepackage{psfig}
\usepackage{epsf}
\usepackage{graphics}
\usepackage{lscape}
\def\edcomment#1{\iffalse\marginpar{\raggedright\sl#1\/}\else\relax\fi}
\reversemarginpar

\begin{document}
\title{Searching for Progenitors of Core-Collapse Supernovae}
\author{Schuyler D.~Van Dyk}
\affil{Spitzer Science Center, 220-6, Pasadena, CA 91125 USA}

\begin{abstract}
Identifying the massive progenitor stars that give rise to core-collapse 
supernovae is one of the main pursuits of supernova and stellar 
evolution studies.  In this talk I discuss some aspects of the pursuit
of these progenitor stars in ground-based and {\sl Hubble Space Telescope\/}
images.
\end{abstract}
\thispagestyle{plain}

\section{Introduction}

The main obstacle to identifying the progenitor of a core-collapse
($M_{\rm ZAMS} \ga 8$--10 $M_{\odot}$) supernova (SN)
is that the SN leaves few traces of the star that exploded.  
Before now, only five out of (at the time of this writing) nearly 3000
historical extragalactic SNe have had their progenitors directly identified.
These include SN 1961V in NGC 1058 (Zwicky 1964, 1965), SN 1978K in NGC 1313 
(Ryder et al.~1993), SN 1987A in the LMC (e.g., Gilmozzi et al.~1987; 
Sonneborn, Altner, \& Kirshner 1987), 
SN 1993J in M81 (Aldering, Humphreys, \& Richmond 1994; Cohen, Darling, 
\& Porter 1995), and SN 1997bs in M66 (Van Dyk et al.~1999).  
It should be noted that these five SNe were all at least somewhat unusual, 
and both SNe 1961V (Van Dyk, Filippenko, \& Li 2002) and 1997bs (Van Dyk et 
al.~2000) may not have been actual SNe.

\section{SN Progenitor Search}

Clearly, direct identification of the progenitors of additional core-collapse
SNe is essential.  Ideally, one pinpoints the exact SN location by comparing 
a late-time SN image with a pre-SN image.  We got the ball rolling for this whole game
by exploiting  the superior
spatial resolution afforded by the {\sl Hubble Space Telescope\/} ({\sl HST})
and identifying the SN 1997bs progenitor, by comparing archival WFPC2 images used
to measure the Cepheid distance to the host galaxy, in which the SN was clearly
detected at late times, with an archival pre-SN F606W WFPC2 image (Van Dyk et al.~1999).

We undertook a more extensive search to isolate the progenitors
of 6 SNe II and 10 SNe Ib/c in WFPC2 images in Van Dyk, Li, \& Filippenko (2003a).
In that paper we recovered SNe 1999dn, 2000C,
and 2000ew at late times, but, unfortunately, their pre-SN images 
did not show a progenitor candidate at the SN position.  For the other 13 SNe
the trick was determining the SN location on one of the four WFPC2 chips.
To achieve the highest astrometric accuracy possible for all image data we had to measure
the SN position on a KAIT image (see the contribution by Filippenko for more on
KAIT)  and then locate the SN site by applying an 
{\it independent\/} astrometric grid to the pre-SN image.  We 
adopted 2MASS (with positional uncertainty
${\la}0{\farcs}10$) as the basis for the grid for both the ground-based 
and {\sl HST\/} images.
Once the SN site was located, photometry of the appropriate WFPC2 chip was
performed using the routine HSTphot (Dolphin 2000a,b) with a 3$\sigma$ detection
threshold.

We had possibly identified the progenitors of the SNe II 1999br, 1999ev,
and 2001du as supergiant stars with $M^0_V \approx -6$ mag, and
the progenitors of the SNe Ib 2001B and 2001is as
very luminous supergiants with $M^0_V \approx -8$ to $-9$ mag, as well as 
the progenitor of the SN Ic 1999bu as a supergiant with $M^0_V\approx -7.5$ mag.
For all other SNe in our sample we could only place limits on the progenitor 
absolute magnitude and color.  

Six of the SNe (1999an, 1999br, 1999ev, 2000ds, 2000ew, 2001B)
had been imaged in multiple bands with ACS at late times by Smartt
and collaborators.  These data became available in the {\sl HST\/} archive.
We had already recovered SN 2000ew, and, along with SNe 1999an and 2000ds, 
we did not detect a SN progenitor in each pre-SN image.
We also found we had not correctly identified the progenitor for SN 2001B (the progenitor,
in fact, is not detected in the pre-SN image).
In Figures 1 and 2 we show that the limits which we are able to place on the progenitors
of the SN Ic 2000ew and the SN Ib 2001B are not very restrictive.
Possibly the SN Ib/c progenitors instead are massive interacting He star binaries (e.g., Avila-Reese 1993), and the theoretical single-star evolutionary tracks shown are not relevant.

\begin{figure}[!ht]
\plotfiddle{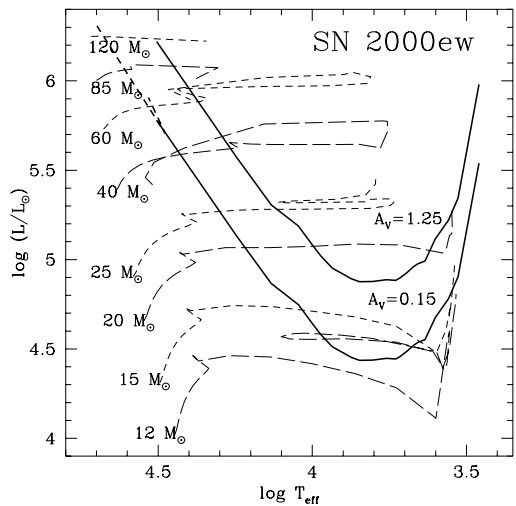}{2.3in}{0}{35}{35}{-115}{-60}
\caption{H-R diagram showing the upper limit ({\it heavy solid line}) to the luminosity, based on
the $V$ upper detection limit, 
of the supergiant progenitor star of the SN Ic 2000ew for a range of possible surface
temperatures (Drilling \& Landolt 2000), and for
two possible extinctions to the SN, $A_V=1.25$ and 0.15 mag (see Van Dyk et al.~2003a).
The {\it heavy dashed line\/} shows the range of luminosity for a possible Wolf-Rayet progenitor star for
the SN.  Stellar evolutionary tracks (alternating {\it long-dashed lines\/} and {\it short-dashed lines}) for a range of initial masses from Lejeune \& Schaerer (2001), with enhanced mass loss and solar
metallicity, are overlaid.}
\end{figure}

\begin{figure}[!ht]
\plotfiddle{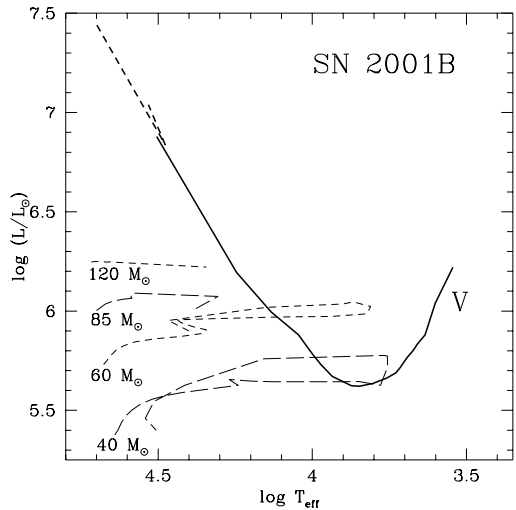}{2.3in}{0}{35}{35}{-115}{-60}
\caption{Same as in Figure 1, based on $V$, but with Galactic foreground $A_V$ (see Van Dyk et al.~2003b), for the progenitor of the SN Ib 2001B.}
\end{figure}

\section{The Type II-P SN 2001du in NGC 1365}

For SN 2001du we isolated three possible candidate progenitor stars
within the uncertainty of the measured SN position: one reddish star with a
detection in both the F555W and F814W bands, and two stars with only F555W
detections, implying bluer colors for these latter two stars (Van Dyk et 
al.~2003a).
Late-time multi-band WFPC2 SN images were obtained by Smartt and collaborators.
We had assumed that, since
SN 2001du is of Type II-P, the most plausible candidate is
the redder star.
We used the late-time SN images, specifically the $V$ image, and
found the SN position on the
pre-SN images to be 0.70$\pm$0.15 WF pixel ($0{\farcs}07$) northeast of one of
the {\it blue\/} stars (Van Dyk, Li, \& Filippenko 2003b).  Thus, we concluded 
that the progenitor is not detected in the pre-SN images.  

We used the F555W and F814W pre-SN image detection limits to
constrain the nature of the progenitor:  Adopting
$E(B-V) \approx 0.1$ mag and distance
modulus $\mu=31.3$ mag for SN 2001du, $M_V>-6.4$ and $M_I>-7.0$ mag for the 
progenitor star.  These limits were converted to the likely
supergiant progenitor luminosity, assuming the full range of possible 
stellar surface temperatures (Drilling \& Landolt 2000).  
Stars with luminosities brighter than these
limits should have been detected in the pre-SN images.
We compared the limits to model stellar evolutionary tracks for a metallicity 
appropriate for the SN environment and for a range of 
masses, and estimated that the SN progenitor mass is 
$M_{\rm ZAMS} < 13^{+7}_{-4}\ M_{\odot}$, which is consistent with the mass
limits on other previous SNe II-P.  See Smartt et al.~(2003) for a
similar result for SN 2001du.  Also see the estimate for the SN 2004dj
progenitor by Maiz-Apellaniz et al.~(2004).

\section{The Type II-P SN 2003gd in Messier 74 (NGC 628)}

Using a precise SN position from ground-based images, we 
isolated the SN position on pre-SN archival WFPC2 images to $\pm 0{\farcs}6$ 
($\pm$6 WF pixels).
Two stars, A and B, were detected near or within the error circle.  
Color information for both progenitor candidates were obtained
from a high-quality, ground-based $I$-band image, on which two faint objects are 
seen near the positions of both A and B.  
Both stars are red supergiants, and from model evolutionary 
tracks for above-solar metallicity, assuming $E(B-V) = 0.13$ mag and
$\mu=29.3$ mag for SN 2003gd, Star B had initial mass 
$M_{\rm ZAMS} \approx 5\ M_{\odot}$ (formally
below the theoretical lower limit for core-collapse SNe), and
Star A had $M_{\rm ZAMS} \approx \ 8$--9 $M_{\odot}$.  
Although Star A is farther from the SN position we measured than is Star B, 
and just outside the edge of the error circle, Star A was considered the most 
plausible progenitor candidate, based on its initial mass and the fact that it
was the brightest $I$-band object within the SN's larger, $\sim$1{\arcsec} radius, 
environment (Van Dyk, Li, \& Filippenko 2003c).  This identification was 
confirmed via late-time {\sl HST\/} imaging of the SN by Smartt et al.~(2004),
who arrive at the same conclusion for the progenitor's initial mass.  (See
the contributions by Hendry and Maund to these proceedings.)

\section{SN 1999br:  A Massive, Failed SN?}

Nomoto et al.~(2004) argued that the faint SNe II-P 1997D and 1999br (e.g., Zampieri
et al.~2003) are highly
massive ($M_{\rm ZAMS} \sim 20$--30 $M_{\odot}$), ``failed'' SNe (presumably a massive envelope
smothers the core-collapse energy release).  To the contrary, when the luminosity of
the possible progenitor of SN 1999br progenitor seen in {\sl HST\/} images (Van Dyk et 
al.~2003a) is placed in context with stellar evolution models (Figure 3), and the 
progenitor is assumed to be a red supergiant, the upper mass limit is
$M_{\rm ZAMS} \sim$12--15 $M_{\odot}$.  (A similar result is found even if the candidate star we 
identified was {\it not\/} the progenitor, and we can only place an upper limit on the 
progenitor luminosity.)

\begin{figure}[!ht]
\plotfiddle{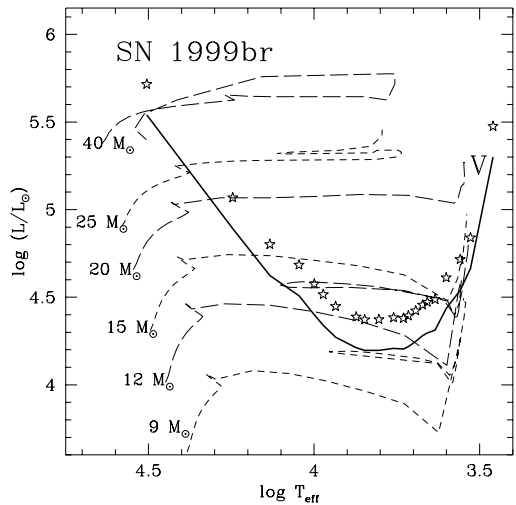}{2.3in}{0}{35}{35}{-115}{-60}
\caption{Same as in Figure 1, based on $V$, with Galactic foreground $A_V$, 
for the progenitor candidate ({\it five-pointed stars})
of the SN II-P 1999br (see Van Dyk et al.~2003a).
Also shown is the upper limit to the progenitor's luminosity ({\it heavy solid curve}), 
if the star identified by us was not the progenitor.  See text for discussion.}
\end{figure}

\section{Progenitors of SN Impostors}

As mentioned above, a growing number of SNe have been shown to likely not
be SNe at all, but are more likely analogs of $\eta$ Car.  These so-called 
``SN impostors'' include SNe 1961V and 1997bs, but
also SN 2002kg, among others.  We (Van Dyk et al.~1999; see our Figure 7a) 
discovered the SN 1997bs progenitor in archival 
F606W images of NGC 3627 from 1994 Dec, at $m_{\rm F606W}=22.86$
mag.  For the distance modulus to the host and a
revised estimate of extinction, we (Van Dyk et al.~2000) found an absolute
magnitude $M_V \simeq -8.1$, which is consistent with it having been an 
extremely luminous supergiant star.  Unfortunately, no color information exists for 
this star.  We can constrain its initial mass by comparing this luminosity and
a range of possible supergiant surface temperatures (Drilling \& Landolt 2000)
to stellar evolutionary tracks; see Figure 4.  The minimum mass 
allowed by the luminosity, for $T_{\rm eff}\approx 6300$ K, is $M\approx 20\ M_{\sun}$.
If the precursor was a blue supergiant, however, the mass can range up to 
120 $M_{\sun}$ or more.

\begin{figure}[!ht]
\plotfiddle{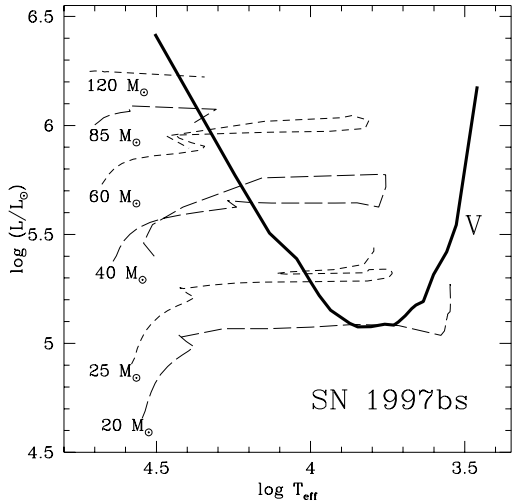}{2.3in}{0}{35}{35}{-110}{-60}
\caption{H-R diagram showing model stellar evolutionary tracks
(alternating {\it long-dashed lines\/} and {\it short-dashed lines}) for a
range of initial masses from Lejeune \& Schaerer (2001), 
with enhanced mass loss for
the most massive stars and solar metallicity.  Also shown is the precursor
luminosity ({\it heavy solid line}) at F606W (${\sim}V$) in the pre-SN {\sl HST\/} 
image (see Van Dyk et al.~1999) for a range of supergiant star surface
temperatures (Drilling \& Landolt 2000).}
\end{figure}

\noindent
The precursor of SN 2002kg has been identified in high quality, multi-band
ground-based, pre-outburst images of the host galaxy.
The star had been previously identified by Tammann \& Sandage (1968) as the
``irregular blue variable'' V37, which had an erratic $B$ light curve over
several decades.
The star had ${M_V}_0=-7.4$ mag and unreddened colors consistent with a very
luminous OB supergiant (the spectra at outburst resemble those of known
B-type LBVs).  Placing the star on a H-R diagram we conclude
that $M_{\rm ZAMS} \ga 60\ M_{\sun}$ for the star (unfortunately, the photometric
uncertainties do not allow this mass to be more tightly constrained).  This is the first
time that an $\eta$ Car analog has had its initial mass accurately estimated.

\section{Conclusions}

We are continuing our search for core-collapse SN progenitors in high-quality
(mainly, {\sl HST}) images.
The possible detections and constraints on the SN II progenitors are broadly
consistent with red supergiants as progenitor stars, with their colors 
implying spectral types typically M or somewhat earlier.  In fact, SNe II-P progenitors appear to
have $M_{\rm ZAMS} \la 20\ M_{\sun}$.  The SN II-P 2003gd progenitor is only the sixth 
ever directly identified, with $M_{\rm ZAMS} =8$--$9\ M_{\sun}$.  The data so far for
the progenitors of SNe Ib/c are inadequate; we are unable to place rigorous constraints 
on either the Wolf-Rayet star or massive interacting binary models for the
progenitors of these SNe Ib/c, the most extreme examples of which have been
associated with some GRBs (see Matheson, these proceedings).  Direct identification of
the progenitors of SNe Ib/c is of the utmost importance, given this connection.
The SNe IIn are a very heterogeneous group, a topic that deserves considerably more
discussion elsewhere.  Some SNe IIn are ``SN impostors,'' which are
more likely superoutbursts of very massive evolved stars, similar to $\eta$ Car.  This
is borne out by the identification of the SN 2002kg progenitor, or precursor, with
$M_{\rm ZAMS} \ga 60\ M_{\sun}$.  If SN 2002kg is instead a real SN (which we doubt),
then its progenitor is only the seventh, out of 1000's, to be directly identified.

\acknowledgments

W.~Li and A.V.~Filippenko are my two co-conspirators in this work.

\end{document}